\documentclass[aps,prl,twocolumn,superscriptaddress,showpacs,preprintnumbers,amsmath,amssymb,floatfix,nofootinbib]{revtex4}

\usepackage{graphicx}
\usepackage{dcolumn}
\usepackage{bm}
\usepackage{url} 
\usepackage{amsmath} 
\usepackage{amssymb} 

\begin{document}

\title{\quad\\[1.0cm] 
Evidence for $CP$ Violation in the Decay $D^+\rightarrow K^0_S\pi^+$
}
\affiliation{Budker Institute of Nuclear Physics SB RAS and Novosibirsk State University, Novosibirsk 630090}
\affiliation{Faculty of Mathematics and Physics, Charles University, Prague}
\affiliation{University of Cincinnati, Cincinnati, Ohio 45221}
\affiliation{Department of Physics, Fu Jen Catholic University, Taipei}
\affiliation{Justus-Liebig-Universit\"at Gie\ss{}en, Gie\ss{}en}
\affiliation{Hanyang University, Seoul}
\affiliation{University of Hawaii, Honolulu, Hawaii 96822}
\affiliation{High Energy Accelerator Research Organization (KEK), Tsukuba}
\affiliation{Hiroshima Institute of Technology, Hiroshima}
\affiliation{Indian Institute of Technology Guwahati, Guwahati}
\affiliation{Indian Institute of Technology Madras, Madras}
\affiliation{Institute of High Energy Physics, Chinese Academy of Sciences, Beijing}
\affiliation{Institute of High Energy Physics, Vienna}
\affiliation{Institute of High Energy Physics, Protvino}
\affiliation{Institute for Theoretical and Experimental Physics, Moscow}
\affiliation{J. Stefan Institute, Ljubljana}
\affiliation{Kanagawa University, Yokohama}
\affiliation{Institut f\"ur Experimentelle Kernphysik, Karlsruher Institut f\"ur Technologie, Karlsruhe}
\affiliation{Korea Institute of Science and Technology Information, Daejeon}
\affiliation{Korea University, Seoul}
\affiliation{Kyungpook National University, Taegu}
\affiliation{\'Ecole Polytechnique F\'ed\'erale de Lausanne (EPFL), Lausanne}
\affiliation{Faculty of Mathematics and Physics, University of Ljubljana, Ljubljana}
\affiliation{Luther College, Decorah, Iowa 52101}
\affiliation{University of Maribor, Maribor}
\affiliation{Max-Planck-Institut f\"ur Physik, M\"unchen}
\affiliation{University of Melbourne, School of Physics, Victoria 3010}
\affiliation{Graduate School of Science, Nagoya University, Nagoya}
\affiliation{Kobayashi-Maskawa Institute, Nagoya University, Nagoya}
\affiliation{Nara Women's University, Nara}
\affiliation{National Central University, Chung-li}
\affiliation{National United University, Miao Li}
\affiliation{Department of Physics, National Taiwan University, Taipei}
\affiliation{H. Niewodniczanski Institute of Nuclear Physics, Krakow}
\affiliation{Nippon Dental University, Niigata}
\affiliation{Niigata University, Niigata}
\affiliation{University of Nova Gorica, Nova Gorica}
\affiliation{Osaka City University, Osaka}
\affiliation{Pacific Northwest National Laboratory, Richland, Washington 99352}
\affiliation{Research Center for Nuclear Physics, Osaka University, Osaka}
\affiliation{RIKEN BNL Research Center, Upton, New York 11973}
\affiliation{University of Science and Technology of China, Hefei}
\affiliation{Seoul National University, Seoul}
\affiliation{Sungkyunkwan University, Suwon}
\affiliation{School of Physics, University of Sydney, NSW 2006}
\affiliation{Tata Institute of Fundamental Research, Mumbai}
\affiliation{Excellence Cluster Universe, Technische Universit\"at M\"unchen, Garching}
\affiliation{Toho University, Funabashi}
\affiliation{Tohoku Gakuin University, Tagajo}
\affiliation{Tohoku University, Sendai}
\affiliation{Department of Physics, University of Tokyo, Tokyo}
\affiliation{Tokyo Institute of Technology, Tokyo}
\affiliation{Tokyo Metropolitan University, Tokyo}
\affiliation{Tokyo University of Agriculture and Technology, Tokyo}
\affiliation{CNP, Virginia Polytechnic Institute and State University, Blacksburg, Virginia 24061}
\affiliation{Wayne State University, Detroit, Michigan 48202}
\affiliation{Yamagata University, Yamagata}
\affiliation{Yonsei University, Seoul}

  \author{B.~R.~Ko}\affiliation{Korea University, Seoul} 
  \author{E.~Won}\email[Correspoding author~~]{eunil@hep.korea.ac.kr}\affiliation{Korea University, Seoul} 
  \author{I.~Adachi}\affiliation{High Energy Accelerator Research Organization (KEK), Tsukuba} 
  \author{H.~Aihara}\affiliation{Department of Physics, University of Tokyo, Tokyo} 
  \author{D.~M.~Asner}\affiliation{Pacific Northwest National Laboratory, Richland, Washington 99352} 
  \author{V.~Aulchenko}\affiliation{Budker Institute of Nuclear Physics SB RAS and Novosibirsk State University, Novosibirsk 630090} 
  \author{T.~Aushev}\affiliation{Institute for Theoretical and Experimental Physics, Moscow} 
  \author{T.~Aziz}\affiliation{Tata Institute of Fundamental Research, Mumbai} 
  \author{A.~M.~Bakich}\affiliation{School of Physics, University of Sydney, NSW 2006} 
  \author{K.~Belous}\affiliation{Institute of High Energy Physics, Protvino} 
% ...
\collaboration{The Belle Collaboration}
  
\begin{abstract}
We observe evidence for $CP$ violation in the decay $D^+\rightarrow K^0_S\pi^+$
using a data sample with an integrated luminosity of 977 fb$^{-1}$ collected by
the Belle detector at the KEKB $e^+e^-$ asymmetric-energy collider. The $CP$
asymmetry in the decay is measured to be $(-0.363\pm0.094\pm0.067)\%$, which is
3.2 standard deviations away from zero, and is consistent with the expected
$CP$ violation due to the neutral kaon in the final state.
\end{abstract}
\maketitle

{\renewcommand{\thefootnote}{\fnsymbol{footnote}}}
\setcounter{footnote}{0}
In the standard model (SM), violation of the combined charge-conjugation and
parity symmetries $(CP)$ arises from a nonvanishing irreducible phase in the
Cabibbo-Kobayashi-Maskawa flavor-mixing matrix~\cite{KM}. In the SM, $CP$
violation in the charm sector is expected to be very small,
$\mathcal{O}(0.1\%)$ or below~\cite{SMCP_YAY}. Since the discovery of
the $J/\psi$~\cite{JPSI1_JPSI2} and the subsequent discovery of open charm
particles~\cite{OPEND0_OPENDP}, $CP$ violation in charmed particle decays has
been searched for extensively and only recently became experimentally
accessible. To date, after the FOCUS~\cite{FOCUS_KSPI}, CLEO~\cite{CLEO_KSPI},
Belle~\cite{OLDKSPI}, and BaBar~\cite{BABAR_KSPI} measurements, the world
average of the $CP$ asymmetry in the decay $D^+\rightarrow
K^0_S\pi^+$~\cite{CC} is $(-0.54\pm0.14)\%$, which is the first evidence of
$CP$ violation in charmed particles. However, it should be noted that the
observed asymmetry is consistent with that expected due to the neutral kaon in
the final state and is not ascribed to the charm sector. Recently, LHCb
reported $\Delta A_{CP}=(-0.82\pm0.21\pm0.11)\%$, where $\Delta A_{CP}$ is the
$CP$ asymmetry difference between $D^0\rightarrow K^+K^-$ and
$D^0\rightarrow\pi^+\pi^-$ decays~\cite{CPV_LHCB}. This is the first evidence
of non-zero $\Delta A_{CP}$ in charmed particle decays from a single experiment.

In this Letter we report the first evidence for $CP$ violation in charmed meson
decays from a single experiment and in a single decay mode, $D^+\rightarrow
K^0_S\pi^+$, where $K^0_S$ decays to $\pi^+\pi^-$. The $CP$ asymmetry in the
decay, $A_{CP}$, is defined as
\begin{eqnarray}
  \nonumber
  A^{D^+\rightarrow K^0_S\pi^+}_{CP}
  &\equiv&\frac
  {\Gamma(D^+\rightarrow K^0_S\pi^+)-\Gamma(D^-\rightarrow K^0_S\pi^-)}
  {\Gamma(D^+\rightarrow K^0_S\pi^+)+\Gamma(D^-\rightarrow K^0_S\pi^-)}\\
  &=&A^{\Delta C}_{CP} + A^{\bar{K}^0}_{CP},
  \label{EQ:ACP}
\end{eqnarray}
where $\Gamma$ is the partial decay width, and $A^{\Delta C}_{CP}$ and $A^{\bar{K}^0}_{CP}$~\cite{ACPK0B} denote $CP$ asymmetries in the charm decay ($\Delta C$) and in $K^0-\bar{K}^0$ mixing in the SM~\cite{BIGI,AZIMOV_XING}, respectively. The observed $K^0_S\pi^+$ final state is a coherent sum of amplitudes for $D^+\rightarrow\bar{K}^0\pi^+$ and $D^+\rightarrow K^0\pi^+$ decays where the former is Cabibbo-favored (CF) and the latter is doubly Cabibbo-suppressed (DCS). In the absence of direct $CP$ violation in CF and DCS decays (as expected within the SM), the $CP$ asymmetry in $D^+\rightarrow K^0_S\pi^+$ decay within the SM is $A^{\bar{K}^0}_{CP}$, which is measured to be $(-0.332\pm0.006)$\%~\cite{ACPK0BII} from $K^0_L$ semileptonic decays~\cite{PDG2010}. On the other hand, if processes beyond the SM contain additional weak phases other than the one in the Kobayashi-Maskawa ansatz~\cite{KM}, interference between CF and DCS decays could generate an $\mathcal{O}(1\%)$ direct $CP$ asymmetry in the decay $D^+\rightarrow K^0_S\pi^+$~\cite{BIGI}. Thus, observation of $A_{CP}$ inconsistent with $A^{\bar{K}^0}_{CP}$ in $D^+\rightarrow K^0_S\pi^+$ decay would be strong evidence for processes involving new physics~\cite{BIGI,LIPKIN_GAO}.

We determine $A^{D^+\rightarrow K^0_S\pi^+}_{CP}$ by measuring the asymmetry in
the signal yield
\begin{equation}
  A^{D^+\rightarrow K^0_S\pi^+}_{\rm rec}=\frac
  {N_{\rm rec}^{D^+\rightarrow K^0_S\pi^+}-N_{\rm rec}^{D^-\rightarrow K^0_S\pi^-}}
  {N_{\rm rec}^{D^+\rightarrow K^0_S\pi^+}+N_{\rm rec}^{D^-\rightarrow K^0_S\pi^-}},    
  \label{EQ:ARECONI}
\end{equation}
where $N_{\rm rec}$ is the number of reconstructed decays. The asymmetry in
Eq.~(\ref{EQ:ARECONI}) includes the forward-backward asymmetry ($A_{FB}$) due
to $\gamma^{*}$-$Z^0$ interference and higher order QED effects in
$e^+e^-\rightarrow c\bar{c}$~\cite{HIGHQED}, and the detection efficiency
asymmetry between $\pi^+$ and $\pi^-$ ($A^{\pi^+}_{\epsilon}$) as well as
$A_{CP}$. In addition, Ref.~\cite{K0MAT} calculated another source denoted
$A_{\mathcal{D}}$ due to the differences in interactions of $\bar{K}^0$ and
$K^0$ mesons with the material of the detector. (The existence of this effect
was pointed out in Ref.~\cite{OLDKSPI}.) Since we reconstruct the $K^0_S$ with
$\pi^+\pi^-$ combinations, the $\pi^+\pi^-$ detection asymmetry cancels out for
$K^0_S$. The asymmetry of Eq.~(\ref{EQ:ARECONI}) can be written as
\begin{eqnarray}
  \nonumber
  A^{D^+\rightarrow K^0_S\pi^+}_{\rm rec}&=&A^{D^+\rightarrow K^0_S\pi^+}_{CP}~+~A^{D^+}_{FB}(\cos\theta^{\rm CMS}_{D^+})\\
  &+&A^{\pi^+}_{\epsilon}(p^{\rm lab}_{T\pi^+},\cos\theta^{\rm
  lab}_{\pi^+})~+~A_{\mathcal{D}}(p^{\rm lab}_{K^0_S})
  \label{EQ:ARECONII}
\end{eqnarray}
by neglecting the terms involving the product of asymmetries. In
Eq.~(\ref{EQ:ARECONII}), $A_{CP}$ is independent of all kinematic variables
other than $K^0_S$ decay time due to the $K^0_S$ in the final
state~\cite{GROSSMAN_NIR}, $A^{D^+}_{FB}$ is an odd function of the cosine of
the polar angle of the $D^+$ momentum in the center-of-mass system (CMS),
$A^{\pi^+}_{\epsilon}$ depends on the transverse momentum and the polar angle
of the $\pi^+$ in the laboratory frame (lab), and $A_{\mathcal{D}}$ is a
function of the momentum of the $K^0_S$ in the lab. To correct for
$A^{\pi^+}_{\epsilon}$ in Eq.~(\ref{EQ:ARECONII}), we use $D^+\rightarrow
K^-\pi^+\pi^+$ and $D^0\rightarrow K^-\pi^+\pi^0$ decays, and assume the same
$A_{FB}$ for $D^+$ and $D^0$ mesons. Since these are CF decays for which the
direct $CP$ asymmetry is expected to be negligible, in analogy to
Eq.~(\ref{EQ:ARECONII}), $A^{D^+\rightarrow K^-\pi^+\pi^+}_{\rm rec}$ and
$A^{D^0\rightarrow K^-\pi^+\pi^0}_{\rm rec}$ include $A_{FB}$,
$A^{K^-}_{\epsilon}$, and $A^{\pi^+}_{\epsilon}$. Thus with the additional
$A^{\pi^+}_{\epsilon}$ term in $A^{D^+\rightarrow K^-\pi^+\pi^+}_{\rm rec}$,
one can measure $A^{\pi^+}_{\epsilon}$ by subtracting $A^{D^0\rightarrow
  K^-\pi^+\pi^0}_{\rm rec}$ from $A^{D^+\rightarrow K^-\pi^+\pi^+}_{\rm
  rec}$. We obtain $A_{\mathcal{D}}$ according to Ref.~\cite{K0MAT}. Using
$A^{D^+\rightarrow K^0_S\pi^+_{\rm corr}}_{\rm rec}$ shown in
Eq.~(\ref{EQ:ARECCORR}), which is $A^{D^+\rightarrow K^0_S\pi^+}_{\rm rec}$
after the $A^{\pi^+}_{\epsilon}$ and $A_{\mathcal{D}}$ corrections,
\begin{equation} 
  A^{D^+\rightarrow K^0_S\pi^+_{\rm corr}}_{\rm rec}~=~A^{D^+\rightarrow
  K^0_S\pi^+}_{CP}~+~A^{D^+}_{FB}(\cos\theta^{\rm CMS}_{D^+}),
  \label{EQ:ARECCORR}
\end{equation}
we extract $A_{CP}$ and $A_{FB}$ using
\begin{subequations} 
\begin{eqnarray}
  \nonumber
  A^{D^+\rightarrow K^0_S\pi^+}_{CP}&=&[A^{D^+\rightarrow K^0_S\pi^+_{\rm corr}}_{\rm rec}(+\cos\theta^{\rm CMS}_{D^+})\\    
    &+&~A^{D^+\rightarrow K^0_S\pi^+_{\rm corr}}_{\rm rec}(-\cos\theta^{\rm CMS}_{D^+})]/2,\\
  \nonumber
  A^{D^+}_{FB}&=&[A^{D^+\rightarrow K^0_S\pi^+_{\rm corr}}_{\rm rec}(+\cos\theta^{\rm CMS}_{D^+}) \\
    &-&~A^{D^+\rightarrow K^0_S\pi^+_{\rm corr}}_{\rm rec}(-\cos\theta^{\rm CMS}_{D^+})]/2. 
\end{eqnarray}
\end{subequations} 
Note that extracting $A_{CP}$ in Eq.~(\ref{EQ:ARECCORR}) is crucial in Belle
due to the asymmetric detector acceptance in $\cos\theta^{\rm CMS}_{D^+}$.

The data used in this analysis were recorded at the $\Upsilon(nS)$ resonances
$(n=1,2,3,4,5)$ or near the $\Upsilon(4S)$ resonance with the Belle
detector~\cite{BELLE} at the $e^+e^-$ asymmetric-energy collider
KEKB~\cite{KEKB}. The data sample corresponds to an integrated luminosity of
977 fb$^{-1}$.

\begin{figure}[htbp]
  \includegraphics[height=0.5\textwidth,width=0.47\textwidth]{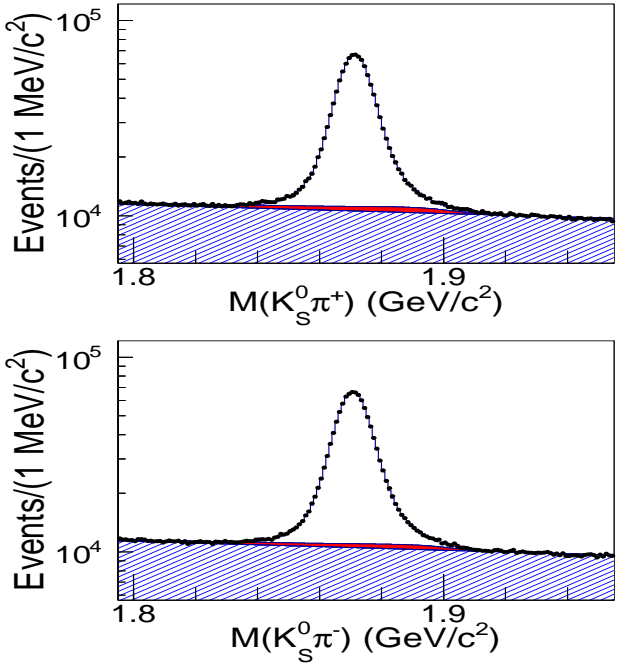}
  \caption{Distributions of $M(K^0_S\pi^+)$ (top) and $M(K^0_S\pi^-)$
  (bottom). Dots with error bars are the data while the histograms show the
  results of the parameterizations of the data. Open histograms represent the
  $D^{\pm}\rightarrow K^0_S\pi^{\pm}$ signal. Shaded and hatched regions are
  $D^{\pm}_s\rightarrow K^0_S K^{\pm}$ misidentification and combinatorial
  backgrounds, respectively.}
  \label{FIG:MKSPI}
\end{figure}

We apply the same charged track selection criteria that were used in
Ref.~\cite{BRKS} without requiring associated hits in the silicon vertex
detector~\cite{SVD2}. We use the standard Belle charged kaon and pion
identification~\cite{BRKS}. We form $K^0_S$ candidates from $\pi^+\pi^-$ pairs,
fitted to a common vertex and requiring the invariant mass of the
pair $M(\pi^+\pi^-)$ to be within $[0.4826, 0.5126]$ GeV/$c^2$, regardless of
whether the candidate satisfies the standard $K^0_S$ requirements~\cite{BRKS}.
(We refer to the $K^0_S$ candidates not satisfying the standard criteria as
``loose $K^0_S$''.)
The $K^0_S$ and $\pi^+$ candidates are combined to form a $D^+$ candidate by
fitting them to a common vertex and the $D^+$ candidate is fitted to the
$e^+e^-$ interaction point to give the production vertex. To remove
combinatorial background as well as $D^+$ mesons, which are produced in
possibly $CP$ violating $B$ meson decays, we require the $D^+$ meson momentum
calculated in the CMS ($p^*_{D^+}$) to be greater than 2.5 and 3.0 GeV/$c$ for
the data taken at the $\Upsilon(4S)$ and $\Upsilon(5S)$ resonances,
respectively. For the data taken below $\Upsilon(4S)$, which is free of $B$
mesons, we apply the requirement $p^*_{D^+}$$>$2.0 GeV/$c$. In addition to the
selections described above, we further optimize the signal sensitivity with
four variables: the $\chi^2$ of the $D^+$ decay and production vertex fits
($\chi^2_D$ and $\chi^2_P$), the transverse momentum of the $\pi^+$
($p_{T\pi^+}$), and the angle between the $D^+$ momentum vector, as
reconstructed from the daughters, and the vector joining the $D^+$ production
and decay vertices ($\xi$)~\cite{DSDCSD}. An optimization is performed by
maximizing $\mathcal{N}_S/\sqrt{\mathcal{N}_S+\mathcal{N}_B}$ with the four
variables varied simultaneously~\cite{ETAH}, where
$\mathcal{N}_S+\mathcal{N}_B$ and $\mathcal{N}_B$ are the yields in the
$K^0_S\pi^+$ invariant mass signal ($[1.855, 1.885]$ GeV/$c^2$) and sideband
($[1.825, 1.840]$ and $[1.900, 1.915]$ GeV/$c^2$) regions, respectively. The
optimal set of ($\chi^2_D$, $\chi^2_P$, $p_{T\pi^+}$, $\xi$) requirements are
found to be ($<$100, $<$10, $>$0.50 GeV/$c$, $<$160$^{\circ}$), ($<$100, $<$10,
$>$0.45 GeV/$c$, $<$170$^{\circ}$), and ($<$100, $<$10, $>$0.40 GeV/$c$, no
requirement) for the data taken below the $\Upsilon(4S)$, at the
$\Upsilon(4S)$, and at the $\Upsilon(5S)$, respectively. The $D^+$ candidates
with the loose $K^0_S$ requirement are further optimized with two additional
variables which are the $\chi^2$ of the fit of pions from the $K^0_S$ decay and
the pion from the $D^+$ meson decay to a single vertex ($\chi^2_{3\pi}$), and
the angle between the $K^0_S$ momentum vector, as reconstructed from the
daughters, and the vector joining the $D^+$ and $K^0_S$ decay vertices
($\zeta$). The two variables are again varied simultaneously and the optimum is
found to be $\chi^2_{3\pi}$$>$6 and $\zeta$$<$4$^{\circ}$ for all data. The
inclusion of $D^+$ candidates with the loose $K^0_S$ requirement improves the
statistical sensitivity by approximately 5\%. After the final selections
described above, there remains a background with a broad peaking structure in
the $K^0_S\pi^+$ invariant mass signal region, due to misidentification of
charged kaons from $D^+_s\rightarrow K^0_S K^+$ decays. The
$D^+\rightarrow\pi^+\pi^-\pi^+$ background is found to be negligible from
simulation~\cite{MC}. Figure~\ref{FIG:MKSPI} shows the distributions of
$M(K^0_S\pi^+)$ and $M(K^0_S\pi^-)$ together with the results of the fits
described below.

The $D^{\pm}\rightarrow K^0_S\pi^{\pm}$ signals are parameterized as a sum of a
Gaussian and a bifurcated Gaussian distribution with a common mean. The
combinatorial background is parameterized with the form $e^{\alpha+\beta
  M(K^0_S\pi^{\pm})}$, where $\alpha$ and $\beta$ are free parameters. The
shapes and normalizations of the $D^{\pm}_s\rightarrow K^0_S K^{\pm}$
misidentification backgrounds are obtained with taking the asymmetry in
$D^{\pm}_s\rightarrow K^0_S K^{\pm}$ into account as described in
Refs.~\cite{BRKS,OLDKSPI}. Both the shapes and the normalizations of the
misidentification backgrounds are fixed in the fit. The asymmetry and the sum
of the $D^+$ and $D^-$ yields are directly obtained from a simultaneous fit to
the $D^+$ and $D^-$ candidate distributions. Besides the asymmetry and the
total signal yield, the common parameters in the simultaneous fit are the
widths of the Gaussian and the bifurcated Gaussian and the ratio of their
amplitudes. The asymmetry and the sum of the $D^+$ and $D^-$ yields from the
fit are $(-0.146\pm0.094)\%$ and (1738$\pm$2)$\times10^3$, respectively, where
the errors are statistical.
\begin{figure}[htbp]
  \includegraphics[height=0.70\textwidth,width=0.47\textwidth]{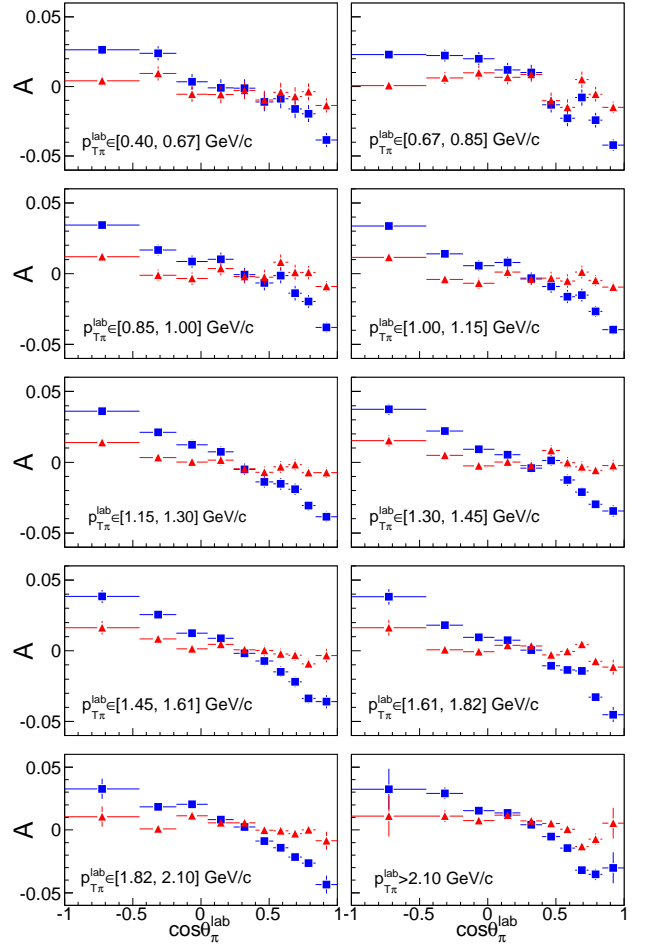}
  \caption{$A^{\pi^+}_{\epsilon}$ map in bins of $p^{\rm lab}_T$ and
  $\cos\theta^{\rm lab}$ of the $\pi^+$ obtained with the $D^+\rightarrow
  K^-\pi^+\pi^+$ and $D^0\rightarrow K^-\pi^+\pi^0$ samples (triangles). The
  $A^{D^+\rightarrow K^-\pi^+\pi^+}_{\rm rec}$ map is also shown (rectangles).}
\label{FIG:APIMAP}
\end{figure}

To obtain $A^{\pi^+}_{\epsilon}$ we first extract $A^{D^0\rightarrow
  K^-\pi^+\pi^0}_{\rm rec}$ from a simultaneous fit with the same
  parameterizations for the signal except for the misidentification
  background. The values of $A^{D^0\rightarrow K^-\pi^+\pi^0}_{\rm rec}$ are
  evaluated in 4$\times$4$\times$4$\times$4$\times$4 bins of the
  five-dimensional (5D) phase space ($p^{\rm lab}_{TK^-}$, $\cos\theta^{\rm
  lab}_{K^-}$, $p^{\rm lab}_{T\pi^+}$, $\cos\theta^{\rm lab}_{\pi^+}$,
  $\cos\theta^{\rm CMS}_{D^0}$). Each $D^{\pm}\rightarrow
  K^{\mp}\pi^{\pm}\pi^{\pm}$ candidate is then weighted with a factor of $1\mp
  A^{D^0\rightarrow K^-\pi^+\pi^0}_{\rm rec}$ in the corresponding bin of the
  5D phase space, where the phase space of the $\pi^{\pm}$ with lower $p_T$ in
  $D^{\pm}\rightarrow K^{\mp}\pi^{\pm}\pi^{\pm}$ decay is used. After this
  weighting, the asymmetry in $D^{+}\rightarrow K^-\pi^+\pi^+$ decay sample
  becomes $A^{\pi^+}_{\epsilon}$, where $\pi^+$ refers to the $\pi^+$ with higher
  $p_T$ in the decay. The detector asymmetry, $A^{\pi^+}_{\epsilon}$, is
  measured from simultaneous fits to the weighted
  $M(K^{\mp}\pi^{\pm}\pi^{\pm})$ distributions in 10$\times$10 bins of the 2D
  phase space ($p^{\rm lab}_{T\pi^+}$, $\cos\theta^{\rm lab}_{\pi^+}$) with the
  same parameterization used in $D^0\rightarrow K^-\pi^+\pi^0$
  decays. Figure~\ref{FIG:APIMAP} shows the measured $A^{\pi^+}_{\epsilon}$ in
  bins of $p^{\rm lab}_{T\pi^+}$ and $\cos\theta^{\rm lab}_{\pi^+}$ together
  with $A^{D^+\rightarrow K^-\pi^+\pi^+}_{\rm rec}$ for comparison. The average
  of $A^{\pi^+}_{\epsilon}$ over phase space is $(+0.078\pm0.040)\%$, where
  the error is statistical.

Based on a recent study of the $A_{\mathcal{D}}$~\cite{K0MAT}, we obtain the
asymmetry in bins of $K^0_S$ momentum in the lab. For the present analysis,
$A_{\mathcal{D}}$ is approximately 0.1\% after integrating over the phase space
of the two-body decay~\cite{K0MAT}.

\begin{figure}[htbp]
  \includegraphics[height=0.4\textwidth,width=0.47\textwidth]{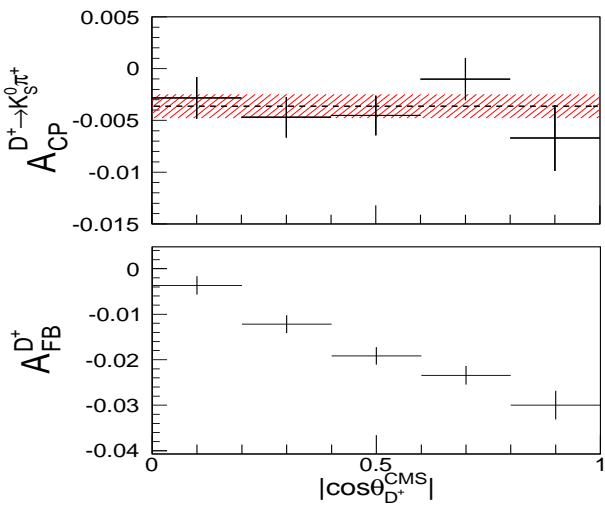}
  \caption{Measured $A_{CP}$ (top) and $A_{FB}$ (bottom) values as a function of
  $|\cos\theta^{\rm CMS}_{D^{+}}|$. In the top plot, the dashed line is the
  mean value of $A_{CP}$ while the hatched band is the $\pm1\sigma_{\rm total}$
  interval, where $\sigma_{\rm total}$ is the total uncertainty.}
\label{FIG:ACP}
\end{figure}

The data samples shown in Fig.~\ref{FIG:MKSPI} are divided into
10$\times$10$\times$16 bins of the 3D phase space ($p^{\rm lab}_{T\pi^+}$,
$\cos\theta^{\rm lab}_{\pi^+}$, $p^{\rm lab}_{K^0_S}$). Each
$D^{\pm}\rightarrow K^0_S\pi^{\pm}$ candidate is then weighted with a factor of
$(1\mp A^{\pi^+}_{\epsilon})(1\mp A_{\mathcal{D}})$ in the 3D phase
space. The weighted $M(K^0_S\pi^{\pm})$ distributions in bins of
$\cos\theta^{\rm CMS}_{D^+}$ are fitted simultaneously to obtain the corrected
asymmetry. We fit the linear component in $\cos\theta^{\rm CMS}_{D^{+}}$ to
determine $A_{FB}$ while the $A_{CP}$ component is uniform in $\cos\theta^{\rm
  CMS}_{D^{+}}$. Figure~\ref{FIG:ACP} shows $A^{D^+\rightarrow
  K^0_S\pi^+}_{CP}$ and $A^{D^{+}}_{FB}$ as a function of $|\cos\theta^{\rm
  CMS}_{D^{+}}|$. From a weighted average over the $|\cos\theta^{\rm
  CMS}_{D^{+}}|$ bins, we obtain $A^{D^+\rightarrow
  K^0_S\pi^+}_{CP}=(-0.363\pm0.094)\%$, where the error is statistical. Without
the $A_{\mathcal{D}}$ correction as in previous
publications~\cite{FOCUS_KSPI,CLEO_KSPI,OLDKSPI,BABAR_KSPI}, the value of
$A_{CP}$ is $(-0.462\pm0.094)\%$.

The method is validated with fully simulated Monte Carlo events~\cite{MC} and
the result is consistent with no input asymmetry.
We also consider other sources of systematic uncertainty. The dominant one in
the $A_{CP}$ measurement is the $A^{\pi^+}_{\epsilon}$ determination, the
uncertainty of which is mainly due to the statistical uncertainties in the
$D^+\rightarrow K^-\pi^+\pi^+$ and $D^0\rightarrow K^-\pi^+\pi^0$
samples. These are found to be 0.040\% and 0.048\%, respectively, from a
simplified simulation study. A possible $A_{CP}$ in the $D^0\rightarrow
K^-\pi^+\pi^0$ final state is estimated with the relation,
$A_{CP}=-y\sin\delta\sin\phi\sqrt{R}$~\cite{PETROV}. Using the 95\% upper and
lower limits on $D^0-\bar{D}^0$ mixing and $CP$ violation
parameters~\cite{HFAG}, $A_{CP}$ in the $D^0\rightarrow K^-\pi^+\pi^0$ final
state is estimated to be less than 0.014\% and this is included as one of
systematic uncertainties in the $A^{\pi^+}_{\epsilon}$ determination. By adding
the contributions in quadrature, the systematic uncertainty in the
$A^{\pi^+}_{\epsilon}$ determination is estimated to be 0.064\%. We estimate
0.003\% and 0.008\% systematic uncertainties due to the choice of the fitting
method and that of the $\cos\theta^{\rm CMS}_{D^+}$ binning, respectively.
Finally, we add the systematic uncertainty in the $A_{\mathcal{D}}$ correction,
which is 0.016\% based on Ref.~\cite{K0MAT}. The quadratic sum of the above
uncertainties, 0.067\%, is taken as the total systematic uncertainty.

We find $A^{D^+\rightarrow K^0_S\pi^+}_{CP}=(-0.363\pm0.094\pm0.067)\%$. This
measurement supersedes our previous determination of $A^{D^+\rightarrow
  K^0_S\pi^+}_{CP}$~\cite{OLDKSPI}. In Table~\ref{TABLE:SUMMARY}, we compare
all the available measurements and give the new world average.
\begin{table}[htbp]
\caption{Summary of $A_{CP}^{D^+\rightarrow K^0_S\pi^+}$ measurements (where the
  first uncertainties are statistical and the second systematic), together with
  their average (where only the total uncertainty is quoted).}
\label{TABLE:SUMMARY}
\begin{ruledtabular}
\begin{tabular}{ll}
Experiment                                & $A^{D^+\rightarrow K^0_S\pi^+}_{CP}$ (\%) \\ \hline
FOCUS~\cite{FOCUS_KSPI}                   & $-$1.6$\pm$1.5$\pm$0.9 \\ 
CLEO~\cite{CLEO_KSPI}                     & $-$1.3$\pm$0.7$\pm$0.3  \\
BaBar~\cite{BABAR_KSPI}                   & $-$0.44$\pm$0.13$\pm$0.10 \\ 
Belle (this measurement)                   & $-$0.363$\pm$0.094$\pm$0.067 \\ \hline
New world average                         &$-$0.41$\pm$0.09  \\ 
\end{tabular}     
\end{ruledtabular}
\end{table}

According to Grossman and Nir~\cite{GROSSMAN_NIR}, we can estimate the
experimentally measured $CP$ asymmetry induced by SM $K^0-\bar{K}^0$ mixing,
$A^{\bar{K}^0}_{CP}$, assuming negligible DCS decay $D^+\rightarrow K^0\pi^+$
in the final state $D^+\rightarrow K^0_S\pi^+$. By multiplying
$A^{\bar{K}^0}_{CP}$ by the correction factor $1.022\pm0.007$ due to the
acceptance effects as a function of $K^0_S$ decay time in our detector, we find
the the measured asymmetry due to the neutral kaons to be $(-0.339\pm0.007)\%$.

In summary, we report evidence for $CP$ violation in the decay $D^+\rightarrow
K^0_S\pi^+$ using a data sample corresponding to an integrated luminosity of
977 fb$^{-1}$ collected with the Belle detector. The $CP$ asymmetry in the
decay is measured to be $(-0.363\pm0.094\pm0.067)\%$, which represents the
first evidence for $CP$ violation in charmed meson decays from a single
experiment and a single decay mode. After subtracting the contribution due to
$K^0-\bar{K}^0$ mixing ($A^{\bar{K}^0}_{CP}$), the $CP$ asymmetry due to the
change of charm ($A^{\Delta C}_{CP}=A^{D^+\rightarrow\bar{K}^0\pi^+}_{CP}$) is
consistent with zero, $A^{\Delta C}_{CP}=(-0.024\pm0.094\pm0.067)\%$. The
measurement in the decay $D^+\rightarrow K^0_S\pi^+$ is the most precise
measurement of $A_{CP}$ in charm decays to date and can be used to place
stringent constraints on new physics models in the charm
sector~\cite{BIGI,LIPKIN_GAO}.

We thank the KEKB group for excellent operation of the accelerator; the KEK
cryogenics group for efficient solenoid operations; and the KEK computer group,
the NII, and PNNL/EMSL for valuable computing and SINET4 network support. We
acknowledge support from MEXT, JSPS and Nagoya's TLPRC (Japan); ARC and DIISR
(Australia); NSFC (China); MSMT (Czechia); DST (India); INFN (Italy); MEST,
NRF, GSDC of KISTI, and WCU (Korea); MNiSW (Poland); MES and RFAAE (Russia);
ARRS (Slovenia); SNSF (Switzerland); NSC and MOE (Taiwan); and DOE and NSF
(USA). B.~R.~Ko acknowledges support by a Korea University Grant, NRF Grant
No. 2011-0025750, and E.~Won by NRF Grant No. 2011-0030865.


\begin{thebibliography}{99}
\bibitem{CC}
Throughout this Letter the charge-conjugate decay modes are implied unless stated otherwise.
\bibitem{ACPK0B}
We use
$A^{\bar{K}^0}_{CP}~=~\frac{\Gamma(\bar{K}^0\rightarrow\bar{f})-\Gamma(K^0\rightarrow
  f)}{\Gamma(\bar{K}^0\rightarrow\bar{f})+\Gamma(K^0\rightarrow f)}$ since
$D^+\rightarrow K^0_S\pi^+$ is dominated by $D^+\rightarrow \bar{K}^0\pi^+$.
\end{thebibliography}
\end{document}